\documentclass[10pt,letterpaper,aps,pra]{revtex4-1}
\usepackage[draft]{hyperref}
\usepackage{braket}
\usepackage{amsfonts}
\usepackage{amsmath}
\usepackage{amssymb}
\usepackage{graphicx}

\begin{document}
\preprint{APS/123-QED}
\title{Quantum properties and dynamics of $X$ states}
\author{Nicol\'as Quesada, Asma Al-Qasimi and Daniel F.V. James}
\affiliation{Department of Physics, University of Toronto, 60 St. George Street, Toronto, Ontario, Canada M5S 1A7}
\email{nquesada@physics.utoronto.ca}

\begin{abstract} 
$X$ states are a broad class of two-qubit density matrices that generalize many states of interest in the literature.
In this work, we give a comprehensive account of various quantum properties of these states, such as entanglement, negativity, quantum discord and other related quantities.
Moreover, we discuss the transformations that preserve their structure both in terms of continuous time evolution 
and discrete quantum processes.
\end{abstract}

\maketitle

\section{Introduction}
The study of the implications of quantum mechanics applied to multipartite correlations has brought many apparent paradoxes \cite{Schrodinger35,Einstein35}  and later many useful applications in information processing\cite{Nielsen2000}, communication\cite{Bennett93}, cryptography\cite{Ekert91} and metrology\cite{Lloyd06}. 
One of the simplest systems that is able to show a lot of the complexities and beautiful subtleties of quantum mechanics is a pair of two level systems (or qubits).
As simple as it may look, many important questions regarding this simple bipartite system have not been or cannot be answered \cite{Girolami11}. While the situation for the description of the correlations of pure quantum states is reasonably well understood, extending the treatment to encompass mixed states, in which quantum and classical correlations both play roles, has led to a veritable menagerie of metrics intended to gauge the ``quantumness'' of a state.  In order to clarify things, in this paper we aim to provide a compendium of the most relevant properties of certain type of two qubits systems known as $X$ states. These states are ubiquitous in the literature as they generalize many important classes of mixed quantum states such as maximally entangled states (like the singlet state or the Bell states), partially entangled and quantum correlated states (like the Werner states), the maximally entangled mixed states \cite{Munro01} as well as non entangled, non quantum-correlated states.
The states that concern us here receive their name for the form of their density matrix:
\begin{eqnarray}\label{def:matrix}
 \rho_X = \left(
\begin{array}{cccc}
 a & 0 & 0 & w \\
 0 & b & z & 0 \\
 0 & z^* & c & 0 \\
 w^* & 0 & 0 & d
\end{array}
\right).
\end{eqnarray}
in the basis $\ket{00}, \ket{01},\ket{10}, \ket{11}$ ($\ket{\alpha \beta} \equiv \ket{\alpha}_A \otimes \ket{\beta}_B$). They have been generated in different physical systems. Two of such systems are the polarizations of a pair of photons generated by a non-linear crystal \cite{Kwiat95} and the electronic levels of a pair of cold ions in a trap \cite{Monroe95}.  
In both cases pure states are generated. Nevertheless, one can imagine that by a desired process\cite{Altepeter03} (such as having the photons pass through a decoherer) or an undesired one \cite{Roos06} (like stray magnetic fields that randomly shift the local energy levels of the cold ions) only coherences between different basis elements are preserved. \\
For instance it might happen that a certain pure two qubit state vector $\ket{\psi}=\alpha \ket{00}+\beta \ket{01}+\gamma \ket{10}+\delta \ket{11}$ is prepared.
These qubits can be encoded in the levels of a pair of trapped cold ions. In this case, the levels will be subject to the Zeeman effect induced by stray magnetic fields and they will suffer a shift 
\begin{equation}
\Delta E = \mu_B g_J m_J B(t)
\end{equation}
where $\mu_B$ is the Bohr magneton, $g_J$ is the gyromagnetic factor, $m_J$ is the magnetic quantum number of the level and $B(t)$ is the fluctuating magnetic field. Under these circumstances the relative phase between the excited ($\ket{1}$) and ground ($\ket{0}$) will be given by:
\begin{equation}
\phi(t)=\frac{\mu_B}{\hbar} \left\{ g_J^{(1)} m_J^{(1)}-g_J^{(0)} m_J^{(0)}   \right\} \int_0^t dt' B(t') dt'
\end{equation}
Typically the fluctuations of the magnetic field will occur in a time scale that is too short to be resolved by any measuring apparatus and thus one will have to resort to a time averaged description \cite{Omar11}. The fluctuations in the phase will induce a decay in certain coherences of the time-averaged density matrix. The evolution of the state vector will be as follows $\ket{\psi(t)}=\alpha \ket{00}+\beta e^{i \phi(t)} \ket{01}+ e^{i \phi(t)}\gamma \ket{10}+\delta e^{i 2\phi(t)} \ket{11}$, but this is just for one realization and one needs consider the average density operator which would be given by:
\begin{eqnarray*}
\bar \rho = \left(
\begin{array}{cccc}
 |\alpha|^2 & \overline{ e^{-i \phi(t) }} \alpha  \beta ^* & \overline{ e^{-i \phi(t) }} \alpha  \gamma ^* &
  \overline{  e^{-2 i \phi(t) }} \alpha  \delta ^* \\
\overline{  e^{i \phi(t) }} \beta  \alpha ^* & |\beta| ^2 & \beta  \gamma ^* & \overline{  e^{-i \phi(t) }}
   \beta  \delta ^* \\
\overline{  e^{i \phi(t) }} \gamma  \alpha ^* & \gamma  \beta ^* & |\gamma| ^2 & \overline{ e^{-i \phi(t) }}
   \gamma  \delta ^* \\
\overline{  e^{2 i \phi(t) }} \delta  \alpha ^* &\overline{  e^{i \phi(t) } }\delta  \beta ^* & \overline{ e^{i \phi(t) }} \delta 
   \gamma ^* & |\delta| ^2
\end{array}
\right), 
\end{eqnarray*}
where the overbar denotes an ensemble average (and the ergodic hypothesis has been invoked to switch from a time to an ensemble average). 
It is reasonable to assume that $\phi$ is normally distributed and thus $\overline{ e^{n i \phi(t) }}=e^{i n \overline{\phi(t)}-n^2 \overline{\phi(t)^2}/2}$.
The magnetic field $B(t)$ can be modelled as white noise with zero mean which implies that $\overline{\phi(t)}=0$ and $\overline{\phi(t)^2}=t/T_2$ where $T_2$ characterizes the variance of the random process $B(t)$.
Under this assumption it is clear that all coherences except the one proportional to $\beta^* \gamma$ approach zero as $t \gg T_2$, and the resultant state is an $X$ state, 
with $a=|\alpha|^2$,  $b=|\beta|^2$,  $c=|\gamma|^2$,  $d=|\delta|^2$,  $a=|\alpha|^2$,  $z=\beta  \gamma ^*$ and $w=0$.  

The above example illustrates one of many possible ways in which an initially arbitrary pure state decoheres into $X$ states. In the following section we intend to investigate general properties of these states. First in section (\ref{def}) we introduce and relate two convenient parameterizations for $X$ states. In section (\ref{corr}) we calculate several important measures of quantum correlations and comment on the type of $X$ states that maximize and minimize such correlations. We also include a measure of classical correlations for these states. Finally, in section (\ref{dyna}) we study the types of dynamics that generate and preserve the shape of $X$ states. 

\section{Definitions and Parameterizations}\label{def}
For a pair of two levels systems, $A$ and $B$, one can define local bases with states that we label $\ket{0}_A, \ket{1}_A$ for system $A$ and $\ket{0}_B, \ket{1}_B$ for system $B$. By taking the tensor product of the basis elements one can construct a basis for the qubit Hilbert space. In such a basis we write a general $X$ state  of the system as:
\begin{eqnarray}\label{def:braket}
 \rho_X& =& a \ket{00} \bra{00}+b \ket{01} \bra{01} + c \ket{10} \bra{10} +d \ket{11} \bra{11} \nonumber \\
&& z \ket{01} \bra{10}+z^* \ket{10}\bra{01}+w \ket{00}\bra{11}+w^* \ket{11}\bra{00}. 
\end{eqnarray}
In the ordered basis $\{\ket{00},\ket{01},\ket{10},\ket{11} \}$ the operator $ \rho_X$ takes the matrix form shown in (\ref{def:matrix}).
Normalized density operators are positive semidefinite and have unit trace, which for the above parameterization implies the following constraints:
\begin{eqnarray}\label{constraints}
a+b+c+d=1, \nonumber \\
a,b,c,d \geq 0, \nonumber \\
|z| \leq \sqrt{b c},  \quad \mbox{and} \quad  |w| \leq \sqrt{a d}. 
\end{eqnarray}
It is always possible to apply a pair of local unitary transformations to make all the coefficients in the definitions (\ref{def:braket}) and (\ref{def:matrix}) non-negative. For instance, the phases of $z$ and $w$ can be absorbed into $\ket{0}_A$ and $\ket{0}_B$, \emph{i.e.} one redefines $e^{i \arg(z)} \ket{0}_A$ as $\ket{0_A}$ and $e^{i \arg(w)} \ket{0}_B$ as $\ket{0_B}$.  Because the correlations of a system do not change when local unitaries are applied to the subsystems then such correlations will depend only on the absolute values of the coherences $z$ and $w$ but not on their phases. From now on we assume that such local unitary transformations have been applied and thus all the coefficients in (\ref{def:matrix}) are real and non-negative.

An equivalent and rather useful way of writing the operator (\ref{def:braket}) is by using the Fano parameterization\cite{Fano57}. To this end, one defines the usual one-qubit Pauli operators,
\begin{eqnarray}
 \sigma_1= \sigma_x=\ket{0}\bra{1}+\ket{1}\bra{0} ;&\quad&  \sigma_2= \sigma_y=i \left(\ket{1}\bra{0}-\ket{0}\bra{1}\right); \nonumber\\
 \sigma_3= \sigma_z=\ket{1}\bra{1}-\ket{0}\bra{0}; &\quad&  \sigma_0= \mathbb{I}=\ket{0}\bra{0}+\ket{1}\bra{1}. \nonumber
\end{eqnarray}
With this definitions one can write:
\begin{eqnarray}\label{fano}
\rho_X=\frac{1}{4}\left\{ \mathbb{I} \otimes  \mathbb{I}+A_3  \sigma_3 \otimes \mathbb{I} + B_3 \mathbb{I} \otimes  \sigma_3+ \sum_{i=1}^3 C_i  \sigma_i \otimes  \sigma_i \right \}  .   \nonumber
\end{eqnarray}
The two parameterizations can be related as follows:
\begin{equation}
\begin{array}{rclrclrcl}
A_3&=&(a+b)-(c+d) \quad & B_3&=&(a+c)-(b+d) \quad & \\
C_1&=&2 (z+w) & C_2&=&2(z-w) & C_3&=&(a+d)-(b+c).
\end{array}
\end{equation}
Notice also that from the Fano parameterization the reduced density matrices can be obtained straightforwardly:
\begin{eqnarray}
\label{marginals}
 \rho_A=\text{tr}_B\left( \rho_X\right)=\frac{1}{2}\left\{ \mathbb{I}+A_3  \sigma_3 \right\}, \quad 
 \rho_B=\text{tr}_A\left( \rho_X\right)=\frac{1}{2}\left\{ \mathbb{I}+B_3  \sigma_3 \right\}. 
\end{eqnarray}
Both density matrices are diagonal in the $\ket{0},\ket{1}$ basis.\\
Finally, notice that $X$ states contain as particular instances many important states like the four maximally entangled Bell states, 
\begin{eqnarray}\label{bells}
\ket{\phi_0}=\frac{1}{\sqrt{2}}\left(\ket{00}+\ket{11} \right), \quad \ket{\phi_i}=( \mathbb{I} \otimes  \sigma_i) \ket{\phi_0} 
\end{eqnarray}
the maximally mixed state, mixtures of maximally entangled and maximally mixed states like Werner states
\begin{eqnarray}
\rho_W=(1-\epsilon) \frac{\mathbb{I}}{4}+\epsilon \ket{\phi_i} \bra{\phi_i},
\label{Wern}
\end{eqnarray}
 and all the states that for a given value of their mixedness (or purity) maximize their entanglement \cite{Munro01}.

\section{Quantum and Classical Correlations}\label{corr}
In this section we provide expressions for several quantum correlations for two-qubit $X$ states. We also include an expression for the classical correlations of these states.
Before calculating any relevant measure of correlations between the two parties in the bipartite system we write its state in two canonical forms. The first one is the eigenvalue decomposition $ \rho=\sum_i \lambda _i \ket{\lambda_i} \bra{\lambda_i}$ with:
\begin{eqnarray}\label{ev}
\lambda_{1/2}&=&u_+ \pm \sqrt{u_-^2+w^2} \\
\lambda_{3/4}&=&r_+ \pm \sqrt{r_-^2+z^2} \nonumber\\
\ket{\lambda_{1/2}} &=& \frac{1}{N_{1/2}}\left(\left\{u_- \pm \sqrt{u_-^2+w^2} \right\}\ket{00}+w\ket{11} \right) \nonumber\\
\ket{\lambda_{3/4}} &=& \frac{1}{M_{1/2}}\left(\left\{r_- \pm \sqrt{r_-^2+z^2} \right\}\ket{01}+z\ket{10} \right) \nonumber
\end{eqnarray}
where $u_{\pm}=(a\pm d)/2$, $r_{\pm}=(b\pm c)/2$ and $N_{1/2}$ and $M_{1/2}$ are normalization constants. Notice that (\ref{constraints}) ensures that all the $\lambda_i$ are positive.
From the above it is straightforward to obtain the von-Neumann entropy as $S( \rho_X)=-\sum_i \lambda_i \log(\lambda_i)$.\\
Using the eigenvalues of $\rho_X$ or directly calculating the trace of the square of the density matrix one obtains the purity of the state:
\begin{eqnarray}
\text{tr}{\rho_X^2}=\sum_i \lambda_i^2=a^2+b^2+c^2+d^2+2w^2+2z^2. \nonumber
\end{eqnarray}
The purity will only be equal to one if either $a=d=w=0$ and $b c =z^2$ or $b=c=z=0$ and $a d =w^2$.\\
From the definitions in this section, we calculate several relevant quantum correlations.\\

\subsection{Entanglement}
For decades one correlation, Quantum Entanglement, was considered the 
defining property that distinguishes quantum systems from classical ones. In simple terms, it was considered to be present in 
a system by the inability to factor its state into a product of the states of the 
subsystems that make it. To be more specific in quantifying entanglement, in the case of 
pure states of two qubits, this is achieved by calculating the von Neumann entropy of the 
reduced density matrix of the system. For mixed states, this method cannot be applied 
directly. First, the mixed state is written in a pure-state decomposition as follows:
\begin{equation}
{ \rho}=\sum_{i}p_{i}|\psi_{i}\rangle\langle \psi_{i}|.
\label{G}
\end{equation}
Entanglement for this mixed state is then defined in terms of the entanglement of the 
pure states involved in the decomposition \emph{and} minimized over all decompositions (since decomposition (\ref{G}) is by no means unique):
\begin{equation}
E({ \rho})=\min\sum_{i} p_{i} E(\psi_{i}),
\label{H}
\end{equation}
Standing on this concept, Wootters defined the famous entanglement measure known as 
concurrence\cite{Wootters98} . In the case of our $X$ states, which in the most general cases 
are mixed, concurrence is given by:
\begin{equation}\label{conc}
\mathcal{C}( \rho_X)=2 \max\left\{0,z-\sqrt{a d},w-\sqrt{b c} \right\}
\end{equation}
Note that this quantity takes values between 0 and 1, the former corresponding to no 
entanglement in the system, the latter to maximal entanglement, and anything in between 
correspond to cases of partially entangled states.

\subsection{Partial Transpose and Negativity}
One of the most powerful techniques for entanglement detection is the use of positive but not completely positive (PNCP) maps. A positive linear map $\Lambda$ between the space of operators acting in two Hilbert spaces $\mathcal{H}_A$ and $\mathcal{H}_B$ satisfies\cite{Guhne2009,bruss2002}:
\begin{eqnarray*}
\text{If} \ \Lambda(X)=Y \ \text{then} \ \Lambda(X^\dagger)=Y^\dagger \\
\text{If} \ X \geq 0 \ \text{then} \ \Lambda(X)=Y \geq 0
\end{eqnarray*}
A positive map $\Lambda$ is completely positive if for an arbitrary Hilbert space $\mathcal{H}_C$ the map $\mathcal{I}_C \otimes \Lambda$ is positive, otherwise is termed positive but not completely positive. Notice that a PNCP map will always map separable density operators to separable density operators. A failure to obtain a positive operator after applying a PNCP over a density operator will imply that the density operator that is given is entangled.
One very useful example of a PNCP map is the transpose. 
In \cite{Peres} the partial transposition of a density operator with respect to one of its subsystems (which is a PNCP) is used to give a very powerful condition for entanglement detection. In \cite{Horo96} it is shown that this condition is necessary and sufficient for entanglement in qubit-qubit and qubit-qutrit systems.
Finally, in \cite{VidalWerner}, this concept is used to quantify entanglement. A partial transpose is performed over the first subsystem (system A here), and the measure called Negativity, $\mathcal{N}(\rho)$, is the sum of the absolute values of the negative eigenvalues of the partially transposed matrix. For arbitrary dimensions if  $\mathcal{N}(\rho)> 0$ there is entanglement in the system, but $\mathcal{N}(\rho)= 0$ gives us no definite answer as to whether the system in entangled or separable in the general case. Nevertheless, as already mentioned for qubit-qubit and qubit-qutrit systems, $\mathcal{N}(\rho)=0$ necessarily implies that $\rho$ is separable.
The partial transposition amounts to swapping the second set of labels in each bra and ket in equation (\ref{def:braket}) which for $X$ states amounts to exchange $z$ and $w$. Thus the conditions for having a positive partially transposed $X$ state are 
 given \cite{Ali09}:
\begin{eqnarray}
z\leq\sqrt{ad} \text{ and } w\leq\sqrt{bc} 
\end{eqnarray}
which is precisely what the concurrence tells us (\ref{conc}) since if any of these conditions is violated the partially transposed $X$ state will not be positive or equivalently the state itself will be entangled. 
As for the precise eigenvalues of the partial transpose of an $X$ states these are given by equation (\ref{ev}) but again doing the change $w \longleftrightarrow z$. In this case only the cases with minus signs before the inequality might give negative eigenvalues and thus the negativity is given by:
\begin{equation*}
\mathcal{N}(\rho)=-\min\left\{0,u_+ - \sqrt{u_-^2+z^2} ,r_+ - \sqrt{r_-^2+w^2}  \right\}
\end{equation*}
where $u_\pm$ and $r_\pm$ are the same quantities that appear in (\ref{ev}).\\
Another interesting property of a partially transposed qubit-qubit density matrix is that it will have at most one negative eigenvalue, and thus its determinant gives a necessary and sufficient condition for entanglement. In \cite{Horo08} this is shown with complete generality, here we show it for $X$ states by elementary means. Notice that if for instance $z \geq \sqrt{a d}$ then from (\ref{constraints}) one has $\sqrt{b c}\geq z$ and $\sqrt{a d} \geq w$ which automatically implies that $\sqrt{b c} \geq w$. The argument is reversed if $w\geq\sqrt{bc} $.
\subsection{Fully Entangled Fraction}
The fully entangled fraction (FEF) is defined as the maximum fidelity that a given quantum state has with a maximally entangled state\cite{Bennett96}. The fidelity between a (generally) mixed state and a pure state is defined as\cite{Jozsa94}:
\begin{eqnarray}
\mathcal{F}( \rho,\ket{\psi}\bra{\psi})=\bra{\psi} \rho\ket{\psi}
\end{eqnarray}
For the fully entangled fraction one calculates the above expectation value with the four Bell states (\ref{bells}).
For $\rho_X$ one has:
\begin{equation}\label{FEF}
\mathcal{E}(\rho_X)=\max(a+d+2w-1,b+c+2z-1).
\end{equation}
For a state with a fixed value of the FEF $\mathcal{E}$ it is easily found that 
the concurrence of such state is bounded by:
\begin{equation}
\mathcal{E} \leq \mathcal{C} \leq \frac{\mathcal{E}+1}{2}
\end{equation}
In particular the states that saturate the upper and lower bounds of the inequality are of the $X$ type\cite{Grond02}. 

\subsection{The Schmidt Number of the State}
An important characteristic of a bipartite quantum state is its ability to encode information about a local quantum process that acts on only one of the qubits. In particular, one asks the question of whether by knowing an initial state $\rho_{\text{in}} $ and the state that is obtained by applying an unknown quantum process in one (say the second) qubit, $\rho_{\text{out}}=\left(\mathcal{I} \otimes \mathcal{L}\right)\left\{\rho_{\text{in}} \right\}$, is it possible to know what the process $\mathcal{L}$ was? In \cite{Alte03} it is shown that the necessary and  sufficient condition to provide complete information about the local process $\mathcal{L}$ is that the Schmidt number of the density matrix equals the square of the dimensionality of the system on which the process acts, in our case $2^2=4$. 

The Schmidt number, which is a familiar concept for pure states, can be found for bi-partite density operators as follows.  For any state $\rho$ one can define the matrix $\Gamma_{\mu,\nu} = {\rm Tr}\{\rho(\sigma_{\mu}\otimes\sigma_{\nu})\}/2$\footnote{The factor of $1/2$ is inserted because the Pauli matrices are orthogonal with respect to the Hilbert-Schmidt inner product $\text{tr}(\sigma_\mu\sigma_\nu)=2 \delta_{\mu,\nu}$ but are obviously not normalized. To account for this fact each Pauli matrix is multiplied by $1/\sqrt{2}$ which for the tensor product of two of them will give rise to the $1/2$ factor. In particular this normalization will guarantee that any pure separable state will have only one non-zero singular value equal to 1.}, where $\mu$ and $\nu$ take the values $(0,1,2,3)$, where $\sigma_{0}$ is the identity.  We may use the singular value decomposition (SVD) to rewrite $\Gamma_{\mu,\nu}$ in the form $\sum_i s_i f_{\mu}^{(i)} g_{\nu}^{(i)}$, where $s_i$ is a set of non-negative real numbers and $f_{\mu}^{(i)}$ and $g_{\nu}^{(i)}$, ($i=1,2,3,4$) are two sets of orthonormal vectors (i.e. $\sum_\mu f_{\mu}^{(i)}f_{\mu}^{(j)}= \sum_\nu g_{\nu}^{(i)}g_{\nu}^{(j)}=\delta_{i,j}$).  Defining the operators $ F_i= \sum_\mu f_{\mu}^{(i)}\sigma_\mu$ and 
$ G_i=\sum_\nu g_{\nu}^{(i)}\sigma_\nu$, we find $\rho=\sum_i s_i F_i \otimes G_i$, which is the Schmidt decomposition of $\rho$.  The operators $ F_i$ and $ G_i$ are orthonormal, \emph{i.e.}, $\text{tr}( F_i ^\dagger  F_j)=\text{tr}( G_i ^\dagger  G_j)=\delta_{i,j}$ (they are not, however positive, hence they do not represent states).  The number of non-zero $s_i$ is the Schmidt Number of $\rho$.   For $X$ states, one finds the singular values are related to the coefficients appearing in the Fano decomposition as follows:
\begin{eqnarray}\label{svd}
s_1 &=&\frac{C_1}{2},\nonumber\\
s_2 &=&\frac{|C_2|}{2},\nonumber\\
s_{3/4} &=& \frac{\sqrt{1+A_3^2+B_3^2+C_3^2\pm D}}{2 \sqrt{2}}, 
\end{eqnarray}
where $D =\sqrt{\left(1+A_3^2+B_3^2+C_3^2\right)^2-4 \left(C_3-A_3 B_3\right)^2}$.

\subsection{Quantum Discord and Classical Correlations}
As described in an earlier subsection, the quantification of entanglement is done using 
the von Neumann entropy. In short, the idea is to use the fact that randomness is 
introduced into the system when a quantumly correlated particle is ignored (by taking the 
partial trace over it). Another method to quantify the strength of the quantum 
correlations of the system is to use the difference between the effect of measurement on 
a classical system compared to a quantum system; \emph{i.e.}, to use the fact that measurements 
disturb quantum systems, but not classical ones. This is the idea that Ollivier and 
Zurek's  Quantum Discord is based on \cite{Ollivier01}. Like Wootters concurrence, quantum discord is defined for two-qubit systems. 
Labeling the two subsystems by $A$ and $B$, a measurement is performed on $B$. If a disturbance 
is detected, then that indicates the existence of quantum correlations, otherwise it 
implies that they are absent. The disturbance is quantified by using the mutual 
information function, which gives an indication of how much information is shared between 
$A$ and $B$. The difference between the mutual information function before the measurement 
and after measurement defines the discord. However, the set of projectors that are applied on 
$B$ have to be chosen so that they give the maximal value for the measurement-induced 
mutual information function. Notice that there is in general a more complex hierarchy of quantum correlations' quantifiers in which different types of measurement schemes are applied of which the Discord is a particular instance \cite{Lang11}.
Although for concurrence, there is an analytic expression for  $X$ states, 
it is not possible to find an analytic expression for discord for the general $X$ state \cite{Lu11}. In general the problem of the calculation of the discord can be cast into the solution of two transcendental equations as it is shown in \cite{Girolami11}.
However, Luo \cite{Luo08} was able to find one subclass of $X$ states for which an analytical expression can be given. This class has maximally 
mixed marginals (MMM):
\begin{eqnarray}
 \rho_A =  \rho_B=\frac{1}{2}  \mathbb{I}
\end{eqnarray}
which implies from equation (\ref{marginals}) that $A_3=B_3=0$. For the MMM states the discord is given by \cite{Luo08}:
\begin{eqnarray}
Q(\rho_{MMM})&=&\frac{1}{4} \left\{(1-|C_1|-|C_2|-|C_3|) \log (1-|C_1|-|C_2|-|C_3|) \right. \nonumber\\
&&(1-|C_1|+|C_2|+|C_3|) \log (1-|C_1|+|C_2|+|C_3|) \nonumber\\
&&(1+|C_1|-|C_2|+|C_3|) \log (1+|C_1|-|C_2|+|C_3|) \nonumber\\
&&\left. (1+|C_1|+|C_2|-|C_3|) \log (1+|C_1|+|C_2|-|C_3|) \right\} \nonumber\\
&&-\frac{1-C}{2}\log(1-C)-\frac{1+C}{2}\log(1+C)
\end{eqnarray}
with $C=\max\{ |C_1|,|C_2|,|C_3| \}$.
Although, obtaining an analytical expression for the discord has been shown to be impossible for states more complex than the MMM states a great deal of advance can be achieved by characterizing the set of states that have zero discord. This was done in \cite{Vedral10}. Following the convention used in \cite{Vedral10}, we assume that measurements are made on subsystem $A$ instead of subsystem $B$. Then it can be shown that the set of states $\Omega$ that have zero discord is given by \cite{Vedral10}:
\begin{eqnarray}
 \rho_{CL}=\sum_k p_k \ket{\psi_k} \bra{\psi_k} \otimes \rho_k^{(B)}
\end{eqnarray}
where $\{ \ket{\psi_k}\}$ is an orthonormal basis set for subsystem $A$.
Since the set $\Omega$ is known, it is possible to define a geometric measure of discord by simply measuring the distance (square norm in the Hilbert-Schmidt space) of a given state to the closest state in the set $\Omega$. This was done in \cite{Vedral10} and the result for $X$ states is simply:
\begin{eqnarray}
\mathcal{D}_A^{(G)}( \rho_X)&=&\frac{1}{4} \min \left\{C_{1}^2+C_{2}^2,C_{1}^2+C_{3}^2+A_3^2\right\} \\ 
 &=& \frac{1}{2} \min \left\{ 4 \left(w^2+z^2\right),  (a-c)^2+(b-d)^2+2(w+z)^2\right\} \nonumber
\end{eqnarray}
The geometric discord when $B$ is measured is simply obtained by replacing $A_3$ by $B_3$ in the first equality or swapping $b$ and $c$ in the second one. Finally, we also point out that an $X$ state is non-discordant if and only if is fully diagonal, which is equivalent to saying that it only has two non-zero singular values (See (\ref{svd})).

Although for general $X$ states, there does not exist an analytic formula for discord, it has been shown that there exists a set of projectors that will give accurate results \cite{Lu11}. These projectors are labelled by the term maximal-correlation-direction measurement (MCDM). Moreover, in \cite{LFFH}, an expression for discord in the case of $b=c$ is derived. Here, we derive a very similar result, but for the slightly more general case in which $b$ and $c$ are not necessarily equal. Using a related approach to these studies, we find an expression that is a very good approximation to discord:
\begin{eqnarray}
Q(\rho_X)&\approx&S(\rho_{B})-S(\rho_X)+   \min\left\{N_{1},N_{2}\right\},
\label{Qap}
\end{eqnarray}
where $S(\rho_X)$ is the von Neumann entropy of the general $X$ state density matrix, $S(\rho_{B})$ is the von Neumann entropy of the reduced density matrix of the second qubit (labelled B, on which the measurement is made) and,
\begin{eqnarray}
N_{1}&=& H\left(\left[\frac{1}{2}+\frac{1}{2}\sqrt{\left(a-d+b-c\right)^{2}+4(z+w)^{2}}\right]\right) \nonumber\\
N_{2}&=&-a \log_{2}\left[\frac{a}{a+c}\right]-b \log_{2}\left[\frac{b}{b+d}\right] - c \log_{2}\left[\frac{c}{a+c}\right]-d \log_{2}\left[\frac{d}{b+d}\right].
\label{N1N2}
\end{eqnarray}
where $H(y)=-y \ \log_2(y)-(1-y) \ \log_2(1-y)$ is the binary entropy function.
In \cite{AsmaDiscord}, a parameterization, in terms of variables $\theta$ and $\phi$, that is used to find discord is defined. In the language of this reference, $N_{1}$ is found for ${\theta = \pi/4, \phi=0}$, and $N_{2}$ is found for ${\theta = \pi/2, \phi=0}$.
To elaborate on the accuracy of this approximate expression for $Q$, when we analyze the $1 \times 10^{5}$ randomly generated $X$ states (see \cite{AsmaDiscord} for method description), we find the following: 0 \% of the points have error $>  10^{-3}$, 0.001 \% have error $> 10^{-4}$, 31.44 \% have error $>  10^{-5}$, 86.10 \% have error $>  10^{-6}$, and 90.78 \% have error $>  10^{-7}$. This shows that if we are only interested in the accuracy up to the forth decimal place, then this is a very good approximation for discord of an $X$ state.

Notice that equation (\ref{Qap}) also gives an approximate expression for the amount of classical correlations that a given state has. Recall that the quantum discord as defined by Zurek and Ollivier is $Q(\rho)=\textbf{I}(\rho)-\textbf{C}(\rho)$, where $\textbf{I}(\rho)=S(\rho_A)+S(\rho_B)-S(\rho)$ is the mutual information function, and $\textbf{C}(\rho)$ is the measurement-induced mutual information function maximized over all measurements on subsystem B. One can also interpret these quantities as follows: $\textbf{I}(\rho)$ represents the \emph{total} (classical and quantum) correlations present in the system, $Q(\rho)$ represents the \emph{quantum} correlations, and $\textbf{C}(\rho)$ represents the \emph{classical} correlations (See \cite{Vedral03,Luo08}, for example). In the light of this interpretation, we can use $\textbf{C}(\rho)$ that we calculated to obtain the $Q$ in (\ref{Qap}) to represent the classical correlations in the $X$ states. It is approximately given by:
\begin{equation}
\textbf{C}(\rho_X) \approx S(\rho_A)-\min\left\{N_{1},N_{2}\right\},
\end{equation}
\noindent where $S(\rho_{A})$ is the von Neumann entropy of the reduced density matrix of subsystem A, and $N_{1}$ and $N_{2}$ are defined in (\ref{N1N2}).

\subsection{Measurement-Induced Disturbance}

Quantum Discord involves finding a local projector (on subsystem B) that minimizes its value. Moreover, Discord is not symmetric, meaning that the discord of A with respect to B is not the same as that of B with respect to A. That is why one has to think which subsystem is more appropriate to perform the local measurement on in order to define discord. These \emph{deficiencies} inspired Luo to introduce the Measurement-Induced Disturbance or simply MID \cite{MIDLuo}.

Like Discord, MID exploits the fact that measurements of quantum systems disturb them in order to capture quantum correlations. However, there are two differences. Now, the local measurements are performed on \emph{both} subsystems A and B. In addition, MID does not require searching for the optimal set of local projectors. Instead, the chosen projectors are constructed using the eigenvectors of the reduced density matrices of A and B. MID is defined as follows \cite{MIDLuo,MIDDatta}:

\begin{equation}
MID= I(\rho)-I(P(\rho)),
\label{MID}
\end{equation}

\noindent where

\begin{equation}
P(\rho)=\sum^{m}_{i=1}\sum^{n}_{j=1}\left(\Pi^{A}_{i}\otimes\Pi^{B}_{j}\right)\rho\left(\Pi^{A}_{i}\otimes\Pi^{B}_{j}\right),
\label{Pr}
\end{equation}

\noindent and $\Pi^{A}_{i}$ and $\Pi^{B}_{j}$ are projectors constructed using the eigenvectors of the reduced density matrices of systems A and B, respectively. Note that for two-qubit systems, and hence the $X$ states we are focusing on in this paper, $m,n=2$. In fact, for $X$ states, $P(\rho)$ is given by:

\begin{eqnarray}\label{def:matrixPr}
P_{X}(\rho) = \left(
\begin{array}{cccc}
 a & 0 & 0 & 0 \\
 0 & b & 0 & 0 \\
 0 & 0 & c & 0 \\
 0 & 0 & 0 & d
\end{array}
\right),
\end{eqnarray}

\noindent and the MID is given by:

\begin{equation}
MID_{X}=-S(\rho_X)+S(P_{X}(\rho)),
\label{MIDX}
\end{equation}

\noindent where $S(\rho_X)$ is the von Neumann entropy of the $X$-matrix before measurement, and $P_{X}(\rho)$ is the von Neumann entropy of the density matrix after the measurement as given in (\ref{def:matrixPr}).

Note, however, that although MID is easier to calculate than discord, for several cases in which there is no quantum advantage, it predicts maximal quantum correlations \cite{AsmaDQCQ1, MIDX}. The approach taken in \cite{MIDX}, in which Ameliorated MID is introduced, addresses this issue. For the $X$ states, it is also shown that MID is ambiguous for the states classified as MMM. However, a simple calculation reveals that if $\Pi^{A}_{i}$ and $\Pi^{B}_{j}$ are constructed using the eigenvectors of $\sigma_{z}$, then $MID=Q$, where $Q$ is given by eq.(\ref{Qap}). It is also shown in \cite{MIDX} that for these states, MID and discord are the same for the Werner states and for the pure states. This implies that the result in (\ref{MIDX}) can be used in these two cases without overestimation of the strength of quantum correlations. Another related observation\cite{eric2011} in a class of states, also named Werner states, is the equality between discord and MID. Note, however, that the latter are not the same as the Werner states we discuss in this work, which are states that are a combination of the identity and a Bell state (See (\ref{Wern})). In \cite{eric2011}, the Werner states are defined to be those that satisfy: $\rho=U \otimes U \ \rho  \ U^{\dagger} \otimes U^{\dagger}$ for any unitary operator $U$. Nevertheless, states in the two different classes have equal discord and MID.

\section{Dynamics}\label{dyna}
In this section we study the types of dynamics and quantum channels that preserve the shape of an $X$ state. One of the reasons why $X$ states became so popular in the study of the dynamics of quantum correlations is because of the relatively simple form that the concurrence takes for such states. Moreover since in such studies \cite{Yu042,Yu04,Asma08} they stay in the $X$ form for all times, the simple equation (\ref{conc}) is valid throughout their evolution. 
To study what types of dynamics preserve the shape of $X$ states we employ the very elegant algebraic characterization of $X$ states presented in \cite{Rau09}. The key ingredient of the characterization is to notice that the set of operators
\begin{eqnarray}
\mathcal{S}=&&\left\{  \mathbb{I} \otimes  \mathbb{I}, \sigma_3 \otimes  \mathbb{I},  \mathbb{I} \otimes  \sigma_3,  \sigma_1 \otimes  \sigma_1,  \sigma_2 \otimes  \sigma_2,  \sigma_3 \otimes  \sigma_3,  \sigma_2 \otimes  \sigma_3,  \sigma_3 \otimes  \sigma_2 \right\} 
\end{eqnarray}
is closed under multiplication, \emph{i.e.}, the product of two of them will be proportional to another element of the set $\mathcal{S}$. The other 8 operators that complete the 16 element basis for the 2 qubit Hilbert-Schmidt space are simply:
\begin{eqnarray}
\mathcal{S'}=&&\left\{  \mathbb{I} \otimes  \sigma_1,  \mathbb{I} \otimes   \sigma_2,  \sigma_1 \otimes  \mathbb{I}, \sigma_2 \otimes \mathbb{I},  \sigma_1 \otimes  \sigma_3,  \sigma_2 \otimes  \sigma_3,  \sigma_3 \otimes  \sigma_2,  \sigma_3 \otimes  \sigma_1\right\} 
\end{eqnarray}
The product of operators belonging to $\mathcal{S}$ and $\mathcal{S'}$ satisfy the following properties: Let $A, B\in \mathcal{S}$ and $C, D \in \mathcal{S'}$ then $A  C \in \mathcal{S'}$, $A B \in \mathcal{S}$ and $C D \in \mathcal{S}$. 
With this elementary observations in mind we characterize the types of unitary evolution and non-unitary evolution 
that map $X$ states to $X$ states.\\
For the Hamiltonian dynamics given by the von-Neumann equation, $\frac{d}{dt}  \rho=i [ \rho, H]$ it is easily shown that $ \rho$ will remain $X$ shaped if  $ H$ is also $X$ shaped. Equivalently $\rho$ will remain $X$ shaped if and only if $H$ is spanned by $\mathcal{S}$.
For non-unitary evolution we first study the case of continuous time evolution. In this case the dynamics is given by a Markovian master equation in the Lindblad form \cite{petruccione}:
\begin{eqnarray}\label{master}
\frac{d}{dt} \rho&=&i[ \rho, H]+\sum_{n,m = 1}^{N^2-1} h_{n,m}\big( 2 L_n \rho  L_m^\dagger - \rho  L_m^\dagger  L_n- L_m^\dagger  L_n\rho\big), \nonumber
\end{eqnarray}
where $N$ is the dimensionality of the Hilbert space in which $ \rho$ acts, $N=4$ in our case. $ H$ dictates the Hamiltonian dynamics that was studied in the previous paragraph. $ L_n$ are a set of orthonormal operators in the Hilbert-Schmidt space to which $ \rho$ belongs. Notice that there are $N^2$ of such operators, but because one of them can always be chosen to be the identity (which will not cause non-unitary dynamics) the sum can be restricted to $N^2-1$. Finally, $h_{n,m}$ are the complex entries of a positive semi-definite matrix. In the non-unitary part of the master equation the density operator appears multiplied by two different operators, $ L_n$ and $ L_m$. These operators will preserve the $X$ shape if and only if $ L_n,  L_m \in \text{span}\{ \mathcal{S} \}$ or $ L_n,  L_m \in \text{span}\{ \mathcal{S'} \}$. If either the operators $ L_m,  L_n$ contain elements of both $\mathcal{S}$ and $\mathcal{S'}$ or they are spanned by different sets ($\mathcal{S}$ and $\mathcal{S'}$) then there will be products involving two operators from $\mathcal{S}$ and one from $\mathcal{S'}$ which, in general, will be outside the space of $X$ states.
Finally, as for quantum channels these are defined in the operator sum representation as \cite{Nielsen2000}:
\begin{eqnarray}
\rho \rightarrow \rho'=\mathcal{L}\{\rho \}=\sum_i X_i \rho X_i^\dagger
\end{eqnarray} 
with $\sum_i X_i X_i^\dagger= \mathbb{I}$. In this case, the same argument applied to the master equation can be used, that is, $\mathcal{L}$ will map $X$ states to $X$ states if and only if the $X_i$ do not mix operators from $\mathcal{S}$ and $\mathcal{S'}$.
\section{Conclusions}
In this paper, we aimed to present some of the most interesting and useful quantum properties in the literature, calculated for $X$ states. This work served as a reminder of already known results for these states, such as those pertaining to concurrence. We also presented some new results. First, by uncovering the singular values for these states, we have provided a tool to determine whether they can be used in ancilla-assisted state tomography. We also calculated measures of entanglement, other than concurrence, which can be of interest. This includes the Negativity and the Fully Entangled Fraction. Moreover, we derived results about the quantum discord of these states, one with regards to geometric discord and the other an approximate analytic expression for discord. The latter is shown to be very accurate when the last digit to be considered is the fourth one after the decimal place. Since it was shown that an analytic expression for discord of states that fall under the class of $X$ states does not exist, this approximate (analytic) result for discord provides an easy and reliable tool in situations where optimization is neither practical nor necessary. Moreover, the measurement-induced disturbance, is calculated for $X$ states, with the added caution that it can only be used for certain states such as the Werner and pure states, and that for the MMM states, the projectors have to be chosen carefully.  The $X$ state MID turns out to be a simple expression, dependent on the entropy of the total system before and that after the measurement is performed. Finally, we discussed the dynamics that preserves the form of the $X$ states by providing an exhaustive classification of the types of quantum master equations and quantum channels that preserve the shape of an initial $X$ state.

\section*{Acknowledgements}
The authors would like to thank Andrew G. White, Eric Chitambar, and Christian Weedbrook for valuable discussions. This work was funded by NSERC. 

\bibliography{xstates}

\end{document}